\documentclass[a4 paper, 12 pt] {article}

\def\be{\begin{equation}}
\def\ee{\end{equation}}
\def\bdi{\begin{displaymath}}
\def\edi{\end{displaymath}}
\def\br{\begin{eqnarray}}
\def\er{\end{eqnarray}}
\def\no{\nonumber}

\def\tr{{\rm tr}}
\def\RR{{\rm I\kern-.1567em R}}                              
 \def\CC{{\rm C\kern-4.7pt                                    
 \vrule height 7.7pt width 0.4pt depth -0.5pt \phantom {.}}} 
 \def\ZZ{{\sf Z\kern-4.5pt Z}}                                

\begin{document}

\begin{titlepage}
\vspace*{-2 cm}
\noindent

\vskip 3cm
\begin{center}
{\Large\bf Conservation laws in Skyrme-type models }
\vglue 1  true cm

C. Adam$^{a*}$,   J. S\'anchez-Guill\'en$^{a**}$,  
and A. Wereszczy\'nski$^{b\dagger}$
\vspace{1 cm}

\small{ $^{a)}$Departamento de Fisica de Particulas, Universidad
      de Santiago}
      \\
      \small{ and Instituto Galego de Fisica de Altas Enerxias (IGFAE)}
     \\ \small{E-15782 Santiago de Compostela, Spain}
      \\ \small{ $^{b)}$Institute of Physics,  Jagiellonian
     University,}
     \\ \small{ Reymonta 4, 30-059 Krak\'{o}w, Poland}

\medskip
\end{center}

\normalsize
\vskip 0.2cm

\begin{abstract}
The zero curvature representation of Zakharov and Shabat has been generalized
recently to higher dimensions and has been used to construct non-linear
field theories which either are integrable 
or contain integrable submodels. 
The Skyrme model,
for instance, contains an integrable subsector with infinitely many 
conserved currents, and the simplest Skyrmion with baryon number one
belongs to this subsector. Here we use a related method, based
on the geometry of target space, to construct a whole class of theories 
which are either integrable or contain integrable subsectors (where
integrability means the existence of infinitely many conservation 
laws). These models have three-dimensional target space, like the Skyrme
model, and their infinitely many conserved currents turn
out to be
Noether currents of the volume-preserving diffeomorphisms
on target space.   
Specifically for the Skyrme model, we find both a weak and a strong
integrability condition, where  the conserved currents form a subset of the
algebra of volume-preserving diffeomorphisms in both cases, but this
subset is a subalgebra only for the weak integrable submodel. 
\end{abstract}

\vfill

{\footnotesize
$^*$adam@fpaxp1.usc.es

$^{**}$joaquin@fpaxp1.usc.es

$^{\dagger}$wereszczynski@th.if.uj.edu.pl }

\end{titlepage}

\section{Introduction}

Non-linear field theories are important in many fields of physics, with
applications ranging from elementary particle theory to condensed matter 
physics. One feature of these theories which adds to their relevance is the
possibility for the existence of extended static (solitons) or stationary
(Q-balls) solutions.   
On the other hand, non-linear field theories 
are notoriously difficult to analyse,
where the degree of difficulty strongly depends on the dimension of the
base space (space-time) on which the fields are defined. In 1+1 dimensions,
an ample mathematical apparatus has been developed for the analysis of
non-linear theories, among which there are the inverse scattering method,
B\"acklund transformations, or the zero curvature representation for integrable
systems, which generalizes the Lax pair representation of finite-dimensional
integrable systems. Integrability, that is, the existence of infinitely many
conserved quantities, is related to all of these methods, and 
seems to be crucial in the analytical treatment of nonlinear theories, like,
e.g., the explicit construction of solutions. 

In higher dimensions, much less is known about non-linear field theories.
A general concept of integrability has not yet been developed there. One may
have, however, theories which contain an integrable subsector like, e.g., in
the non-linear sigma model in 2+1 dimensions, where the integrable subsector
is formed by the holomorphic solitons of Belavin and Polyakov.
One generalization of the zero curvature representation of Shabat and Zakharov
to higher dimensions has been proposed in \cite{AFSG}, and it was demonstrated 
there that this proposal leads to non-linear field theories which have
either infinitely many conservation laws in the full theory, or which contain
integrable subsectors, defined by some additional constraint equations on the
fields, such that the solutions belonging to this subsectors 
have infinitely many conservation laws. This zero curvature representation,
therefore, realizes the concept of integrability in higher-dimensional
non-linear field theories in a specific and well-defined manner.
These methods also led to the investigation of specific models, and to the
analytic construction of both static and time-dependent solutions. For
models with infinitely many conservation laws (the so-called AFZ model
and related models), static and time-dependent
solutions have been constructed, e.g., in \cite{AFZ1}, \cite{AFZ2}, 
\cite{Fer1}, \cite{Wer1}, and in \cite{Fer2}, \cite{Fer3}, respectively. 
Solutions in integrable subsectors of models which are, themselves, not
integrable, have been constructed, e.g., in 
\cite{Nic}, \cite{Wer2}--\cite{ASGVW}
(the Nicole model and versions thereof) and in \cite{Ward1}, \cite{ASGW-S3}
(diverse models on base space $S^3$).

One well-known theory which contains an integrable subsector is the Skyrme
model \cite{Skyrme},  a non-linear field theory with target space 
SU(2) (or, equivalently, the three-sphere $S^3$). 
In addition, the simplest Skyrmion (i.e., the
simplest soliton of the Skyrme model 
with baryon number equal to one) belongs to this
integrable subsector, see \cite{FSG1}. 

The zero curvature representation of Ref. \cite{AFSG} also originated
some mathematical applications \cite{AGus}--\cite{RAtt}, in addition to
the mentioned construction of infinitely many conserved currents. It is
clear that the specific connection one- and two-forms which have been proposed
in \cite{AFSG} are not the most general ones, and therefore the approach is
not so restrictive.
In fact, many applications have shown  that most of the
new conserved currents
in models and their sectors are Noether currents
and generalizations thereof, i.e., 
they are related to
the geometry and symmetries of the target space (see \cite{BF1}). So a
direct, geometric approach has been succesfully undertaken to find
those currents for models on target space $S^2$, 
corresponding to the  Faddeev--Niemi model and
modifications of it (like the AFZ and Nicole models), \cite{ASG1},
\cite{ASGW-Ab}. 
In \cite{ASGW-Ab} it was also found that for models which
are not themselves integrable, but contain integrable submodel defined by some
constraint, there exist, in fact, weaker constraints (which are easier to
obey) which still lead to infinitely many conservation laws. For the sake
of clarity we will call the original, stronger constraints (as found, e.g. 
for the Skyrme model in \cite{FSG1}, or for the Faddeev--Niemi model in
\cite{SG1}) the ``strong integrability conditions'', whereas we will call
the weaker constraints which still lead to infinitely many conservtion laws
(as found in \cite{ASGW-Ab}, or in this paper) the ``weak integrability
conditions''.

It is the main aim of this paper to perform the geometric analysis and
classification of integrable models and submodels for a general class of
Skyrme type models, that is, for theories with the same field content as
the Skyrme model (i.e., with three-dimensional target space), analogously
to what was done in \cite{ASG1}, \cite{ASGW-Ab} for general models with
two-dimensional target space. We will find that due to the greater complexity
of the three-dimensional target space there are significantly more
possibilities for integrable models and submodels. Specifically for the Skyrme
model we rederive both the original strong integrability condition of
\cite{FSG1} and the weaker integrability condition recently found in
\cite{ASGW-SkCPn}. Further, we find a class of models with infinitely many
target space symmetries for which infinitely many soliton solutions have been
constructed recently \cite{ASGW-SkCPn}.

The paper is organized as follows. In Section 2 we work out some issues
of the Skyrme model which we need in the sequel, especially its properties
related to the $S^3$ geometry of its target space. These results are known,
and some aspects thereof are discussed, e.g., in \cite{Mant1}, 
 \cite{Mak}, \cite{PaGi},  \cite{mansut}, but we find it useful for our
purposes to collect them here and to present them in a geometric formulation. 
In Section 3 we introduce
volume-preserving diffeomorphisms on a three-dimensional manifold and
study its properties as well as some subsets thereof. Further, we
introduce the corresponding Noether currents, when the volume-preserving
diffeomorphisms are realized on the target space of a field theory.
In Section 4 we introduce a general class of field theories with
three-dimensional target space. Then we classify all possibilities 
for the conservation of all or some of the Noether currents of Section 3
either in the full theory or in a submodel defined by additional
integrability conditions. As the number of possibilities to realize this
integrability is quite big, we review the results of our classification in
three tables at the end of Section 4. 

Finally, we want to explain briefly our conventions for partial derivatives.
Derivatives w.r.t. target space variables will be denoted frequently by
subindices, e.g., $(\partial / \partial a) f \equiv f_a$, 
$(\partial / \partial \xi ) f \equiv f_\xi $. If an index notation $X^i$ 
is used for target space variables, then the corresponding derivatives are
sometimes written like $(\partial / \partial X^i ) \equiv \partial_i $
or  $(\partial / \partial X^3 ) \equiv \partial_3 $. Derivatives w.r.t.
base space variables (space-time coordinates) are also frequently written
as subindices, e.g. $(\partial /\partial x^\mu ) u \equiv u_\mu$,
$u^\mu \equiv \eta^{\mu\nu} u_\nu$, when they act on scalar functions
(like $u$, $\xi$, $a$, etc.). There are some vector-like quantities which
carry greek letter (space-time) indices which do not mean derivatives,
namely the currents $J_\mu$ and the canonical four-momenta $\pi_\mu$
and $P_\mu$, but there should not be any confusion (greek letter indices
on scalar functions always mean derivatives). Observe that also the
notation $(\partial / \partial u^\mu ) {\cal L} \equiv {\cal L}_{u^\mu }$
does occur.

\section{Skyrme model geometry}
The Skyrme model Lagrangian is
\be
{\cal L}_{\rm Sk} = \frac{m^2}{2} {\cal L}_2 -  \lambda {\cal L}_4
+ \frac{M^4}{2} {\cal L}_0
\ee
where
\be
{\cal L}_2 =
\tr \left( U^{\dagger}
\partial_{\mu} U U^{\dagger} \partial^{\mu} U \right) ,
\ee
\be
{\cal L}_4 =  \tr \left[ U^{\dagger} \partial_{\mu} U,
U^{\dagger} \partial_{\nu} U \right]^2,
\ee
and
\be
{\cal L}_0 = \tr (U-{\bf 1}) 
\ee
where ${\bf 1}$ is the $2\times 2$ unit matrix.
Here $m$ and $M$ are constants with the dimension of mass and $\lambda$ is 
a dimensionless constant. $U$ is a SU(2)-valued
matrix field 
\br
U &:&  \RR \times \RR^3 \, \to \, {\rm SU(2)} \no \\
 && x^\mu \, \to \, U(x^\mu )
\er
which may be parameterized in the standard manner like
\begin{equation}
U=e^{i\vec \xi \cdot \vec \sigma}. \label{def U}
\end{equation}
Here, $\vec \sigma \sim \sigma_i, \; i=1,2,3$ 
are the Pauli matrices and $\xi_i$ are
real fields.  In the sequel we will use the slightly different 
parametrization 
\be
\xi \equiv |\vec \xi | \, \quad \vec n \equiv \frac{\vec \xi}{\xi}
\ee
such that 
\begin{equation}
U=e^{i\xi \vec n \cdot \vec \sigma} =\cos \xi 
{\bf 1} +i \sin \xi \vec n \cdot \vec 
\sigma \label{def2 U}
\end{equation}
and the complex scalar field $u$, which is related to the unit vector field
$\vec n$ by stereographic projection,
\be
{\vec n} = \frac{1}{ (1+ u\bar u)^2} \, ( u+\bar u , -i ( u-\bar u ) ,  
1-u\bar u ) \; ;
\qquad
u  = \frac{n_1 + i n_2}{1 + n_3}.
\label{stereo}
\ee
In terms of these variables the Skyrme model Lagrangian is given by
\be
 {\cal L}_2 = \xi^\mu \xi_\mu + 4 \sin^2 \xi \, \frac{u^\mu \bar u_\mu}{
(1+u\bar u)^2} ,
\ee
\be
{\cal L}_4 = 16\sin^2 \xi \left( \xi^\mu \xi_\mu \frac{u^\nu \bar u_\nu}{
(1+u\bar u)^2} - \frac{\xi^\mu u_\mu \xi^\nu \bar u_\nu}{
(1+u\bar u)^2} \right)  + 16
 \sin^4 \xi \, \frac{(u^\mu \bar u_\mu )^2 - u_\mu^2 \bar u_\nu^2 }{
(1+u\bar u)^4} 
\ee
and
\be
 {\cal L}_0 =\cos \xi -1 .
\ee
The target space SU(2) is equivalent to $S^3$ as a manifold, and the
$S^3$ geometric aspects of the Skyrme model are especially transparent in
the target space coordinates $(\xi ,u ,\bar u)$. The metric of $S^3$
in these coordinates is
\be \label{met-S3}
ds^2 = d\xi^2 +4\frac{\sin^2 \xi}{(1+u\bar u)^2} du d\bar u
\ee
such that the quadratic part of the Skyrme model, ${\cal L}_2$, is just the
pullback under the map $U$ of this metric, whereas the quartic part,
${\cal L}_4$, is the pullback of an induced metric on area twoforms. Let us be
somewhat more precise on this issue. The metric tensor is
\be
{\bf g}= dX^3 \otimes dX^3 + g (dX^1 \otimes dX^1 +  dX^2 \otimes dX^2 )
\equiv \theta^1 \otimes \theta^1 +\theta^2 \otimes \theta^2 +
\theta^3 \otimes \theta^3
\ee
where 
\be \label{tar-coor}
X^3 \equiv \xi \, \quad X^1 +iX^2 \equiv u
\ee
are the corresponding real coordinates and $\theta^i$ are the
co-frame one-forms,
\be
\theta^3 = d X^3 \, ,\quad \theta^1 =\sqrt{g} dX^1 
\, ,\quad \theta^2 =\sqrt{g} dX^2
\ee
where 
\be
 g=4\frac{\sin^2 \xi}{(1+u\bar u)^2} =4\frac{\sin^2 X^3 }{(1+(X^1)^2 
+ (X^2)^2)^2}
\ee
is the volume density for the metric on $S^3$,
\be
g=\det (g_{ij}) \, ,\quad {\bf g}= g_{ij}dX^i \otimes dX^j .
\ee
The pullback under the map $U$ of the co-frame one-forms is, e.g., for
$\theta^1$,
\be
\theta^1 {}' \equiv U^* (\theta^1 ) =\sqrt{g} X^1_\mu dx^\mu ,
\ee
and the quadratic part of the Skyrme Lagrangian, ${\cal L}_2$, is just 
the sum of the squared norms (length) of these pullbacks in base space 
(Minkowski space $\RR^3 \times \RR$),
\be
{\cal L}_2 =\sum_{i=1}^3 \tilde \eta (\theta^i {}' ,\theta^i {}')
\ee
where
\br
\sum_{i=1}^3 \tilde \eta (\theta^i {}' ,\theta^i {}') &=&
(g(X^1_\mu X^1_\nu + X^2_\mu X^2_\nu )  +X^3_\mu X^3_\nu )
\tilde \eta (dx^\mu ,dx^\nu ) \no \\ &=&
(\eta^{\mu\nu} (g(X^1_\mu X^1_\nu + X^2_\mu X^2_\nu )  +X^3_\mu X^3_\nu ) 
\no \\
& \equiv & g((X^1_\mu )^2 + (X^2_\mu )^2 ) + (X^3_\mu )^2 .
\er
Here $\tilde \eta$ is the metric co-tensor in
Minkowski space,
\be
\tilde \eta = \eta^{\mu\nu} \partial_\mu \otimes \partial_\nu
\ee
where $\eta^{\mu\nu} =\eta_{\mu\nu} ={\rm diag}(1,-1,-1,-1)$.
Further, the co-tensor is evaluated on one-forms via the canonical inner
product
\be
\partial_\mu (dx^\nu) = \delta_\mu^\nu 
\ee
as usual.

For an analogous geometric interpretation of the quartic term ${\cal L}_4$
we introduce the three unit area two-forms
\br
\Omega_1 &=& \theta^2 \wedge \theta^3 = \sqrt{g} dX^2 \wedge dX^3 \no \\
\Omega_2 &=& \theta^3 \wedge \theta^1 = \sqrt{g} dX^3 \wedge dX^1 \no \\
\Omega_3 &=& \theta^1 \wedge \theta^2 = g dX^1 \wedge dX^2
\er 
Their pull-backs under the map $U$ are, e.g., for $\Omega_3$,
\be
\Omega_3 {}' \equiv U^* (\Omega_3) =g X^1_\mu X^2_\nu dx^\mu \wedge dx^\nu
\ee
and the quartic part ${\cal L}_4$ is equal to the sum of the squared lengths
of these three pullbacks,
\be
{\cal L}_4 = \sum_{i=1}^3 |\Omega_i {}'|^2 
\ee
where, e.g.,
\br
|\Omega_3{}'|^2 &=& g^2 X^1_\mu X^2_\nu X^1_\alpha X^2_\beta [
\tilde \eta (dx^\mu ,dx^\alpha ) \tilde \eta (dx^\nu ,dx^\beta )
- \tilde \eta (dx^\mu ,dx^\beta ) \tilde \eta (dx^\nu ,dx^\alpha ) ]
\no \\
&=& g^2 (X^1_\mu X^2_\nu X^{1\mu} X^{2\nu} -
X^1_\mu X^2_\nu X^{1\nu} X^{2\mu})  \no \\
& \equiv & g^2[(X^1_\mu)^2 (X^2_\nu)^2 - (X^1_\mu X^{2\mu})^2]
\er 

The potential term (or pion mass term) ${\cal L}_0$ does not have a similar
interpretation in terms of the $S^3$ geometry of the target space. This
term is, however, sometimes omitted, and we shall treat both the case with and
without this term. 

The geometric structure of the Skyrme model without the potential term
 ${\cal L}_0$ is also reflected in its target space symmetries.  Indeed,
both the quadratic and the quartic term are invariant under the transformation
\be
U\to VUW^\dagger \, , \quad V,W\in {\rm SU(2)} .
\ee
(The pion mass term is only invariant under the diagonal subgroup $V=W$.)
The transformation on $U$ is the same for $(V,W)$ and $(-V,-W)$, therefore
the target space symmetry group is SU(2)$\times$SU(2)/$\ZZ_2$ $\sim $ SO(4).
Further, $SO(4)$ is the isometry group of the three-sphere $S^3$, so the
above geometric identification of the Skyrme Lagrangian leads to the 
assumption that the target space symmetries of the Skyrme model (without pion
mass term) are just the isometries of the $S^3$ target space metric. And
this is indeed the case. For infinitesimal transformations
\be
V= 1+i\vec \alpha \cdot \vec \sigma \, ,\quad 
W= 1+i\vec \beta \cdot \vec \sigma
\ee
and using the parametrization in terms of $u,\bar u, \xi$ for $U$, the action
of these infinitesimal transformations on $u,\bar u, \xi$ may be calculated
to be $u\to u+Y^u$, $\xi \to \xi +Y^\xi$ with
\be
Y^u = \frac{1}{2}\left( \frac{\cos\xi}{\sin\xi}(\gamma_+ -u^2 \gamma_-
-2u \gamma_3 ) +i(\delta_+ -u^2 \delta_- -2u\delta_3 )\right) 
\ee
\be
Y^\xi = \frac{1}{1+u\bar u}[\bar u\gamma_+ +u\gamma_- +(1-u\bar u)\gamma_3 ]
\ee
where
\be
\vec \gamma \equiv \vec \alpha -\vec \beta \, ,\quad 
\vec \delta \equiv \vec \alpha + \vec \beta \, ,\quad
\gamma_\pm \equiv \gamma_1 \pm i\gamma_2 \, ,\quad 
\delta_\pm \equiv \delta_1 \pm i\delta_2
\ee
and $\vec \delta$ parametrizes the diagonal subgroup $V=W$.
The six corresponding vector fields into the directions $\vec \gamma$ and $\vec
\delta$ indeed span the Lie algebra of SO(4) and leave invariant the
metric (\ref{met-S3}) on $S^3$ (i.e., the obey the Killing equation), 
so they generate the isometries of the target space
of the Skyrme model. 

These symmetries define six conserved Noether currents in the Skyrme model.
There exist further Noether currents in the model which are not 
conserved in the full model. They are, however, conserved in a submodel
where the fields   $u,\bar u, \xi$ obey, in addition to the field equations,
the constraints
\begin{equation}
(\partial_{\mu}u)^2=0, \;\;\; \partial^{\mu} \xi \partial_{\mu} u
=0 .\label{old int cond2}
\end{equation}
Then, one can construct two classes of infinitely many conserved
currents \cite{FSG1}, namely
\begin{equation} \label{G-curr}
J_{\mu}^G=ig^{-1}( G_u \bar \pi_{\mu} - 
G_{\bar u} \pi_{\mu}) \label{current1}
\end{equation}
and
\begin{equation}
J_{\mu}^{(H^{(1)},H^{(2)})}=(1+u\bar u)^2 [2\frac{\cos\xi}{\sin\xi} 
(H^{(1)}\bar \pi_{\mu}+H^{(2)}\pi_{\mu})
- \left( H^{(1)}_{\bar u} +
 H^{(2)}_u \right) P_{\mu} ] \label{current2}
\end{equation}
where $G$ is an arbitrary function of $\xi, u, \bar u$ whereas
$H^{(1)},H^{(2)}$ depend only on $u$ and $\bar u$. Further, 
\be
\pi_\mu \equiv {\cal L}_{u^\mu} \, ,\quad P_\mu \equiv {\cal L}_{\xi^\mu}
\ee
are the canonical four-momenta of the Skyrme model. The second class of
currents is not real for $H^{(1)} \not= H^{(2)}$ but an equivalent set of
real currents is easily found to be
\begin{equation} \label{H-curr1}
J_{\mu}^{(H^{(1)})}=(1+u\bar u)^2 [2\frac{\cos\xi}{\sin\xi} 
H^{(1)}(\bar \pi_{\mu}+\pi_{\mu})
- \left( H^{(1)}_{\bar u} +
 H^{(1)}_u \right) P_{\mu} ] \label{current2a}
\end{equation}
\begin{equation} \label{H-curr2}
J_{\mu}^{(H^{(2)})}=i(1+u\bar u)^2 [2\frac{\cos\xi}{\sin\xi} 
H^{(2)}(\bar \pi_{\mu} - \pi_{\mu})
- \left( H^{(2)}_{\bar u} -
 H^{(2)}_u \right) P_{\mu} ] . \label{current2b}
\end{equation}
We shall see in the next section that these currents form, in fact, a subset 
of the Noether currents of the volume-preserving diffeomorphisms on
the target space of the Skyrme model.

Remark 1: the Noether currents of the target space symmetries form, of course, 
a subset of the above Noether currents. More precisely, the Noether currents
of the diagonal subgroup ($\vec \gamma =0$) are of the type (\ref{G-curr}),
whereas the other three Noether currents $(\vec \delta =0$) are of the type
(\ref{H-curr1}), (\ref{H-curr2}).

Remark 2: we will see in the next section that
the charges of the Noether currents of the type (\ref{H-curr1}), 
(\ref{H-curr2}) do not close as a Lie algebra, neither among themselves, nor
together with the charges of the currents (\ref{G-curr}), whereas the
charges of the currents (\ref{G-curr}) do close as a Lie algebra among
themselves. There exist, however, subsets of the currents  (\ref{G-curr}),
(\ref{H-curr1}) and (\ref{H-curr2}) which do form a closed Lie algebra,
like, e.g., the six generators of the target space symmetry group
SO(4) (see Remark 1 above).

\section{Volume preserving diffeomorphisms}

The volume $n$-form on an $n$ dimensional manifold ${\cal M}^n$ is
\be
dV = g(X^i) dX^1 \wedge dX^2 \wedge \ldots dX^n \quad ,\quad i=1 ,\ldots \, n
\ee
where we use capital letters $X^i$ for the coordinates, because the manifold 
will be identified with target space later on. Here $g$ is the volume density.
If ${\cal M}^n$ is a Riemannian manifold with metric $g_{ij}$, then the
volume density $g$ is the square-root of the determinant of the metric
$g_{ij}$. A volume-form preserving
diffeomorphism is a coordinate transformation $X^i \to X'^i (X^i)$ such
that the volume form remains invariant, $g(X'^i) dX'^1 \ldots dX'^n
= g(X^i) dX^1 \ldots dX^n$. For an infinitesimal transformation
\be
X'^i =X^i +\epsilon Y^i (X^j)
\ee
invariance of the volume form requires
\be \label{vol-dif-cond}
\partial_i (g Y^i)\equiv 
\frac{\partial}{\partial X^i } (g Y^i )=0.
\ee
If the manifold (target space) is three-dimensional, $n=3$, then a general
(local) solution to this equation is provided by Darboux's theorem,
\be
Z^i \equiv gY^i = \epsilon^{ijk}\partial_j A \partial_k B
\ee
where $A$, $B$ are arbitrary functions of the $X^i$. 
The functions $A$, $B$ are called Clebsch variables, and their choice is not 
unique (that is, different $A$, $B$ can give rise to the same $Y^i$).
Therefore, a general
vector field generating a volume preserving diffeomorphism reads
\be
{\bf v}^{(Y)} =
Y^i \partial_i = h  \epsilon^{ijk}(\partial_j A) (\partial_k B)
\partial_i \quad ,\quad h\equiv g^{-1} .
\ee 
For later convenience we introduce the vector field components corresponding
to the target space coordinates (\ref{tar-coor}) of Section 2, 
\be
Y^u = Y^1 +iY^2 \, \quad Y^{\bar u} = Y^1 -iY^2 \, ,\quad Y^\xi =Y^3 ,
\ee
such that the above vector field is re-expressed like
\br
{\bf v}^{(Y)} 
&=& Y^u \partial_u +Y^{\bar u} \partial_{\bar u} +Y^\xi \partial_\xi 
\no \\  &=&
2ih [ (A_\xi B_{\bar u} - A_{\bar u} B_\xi ) \partial_u 
- (A_\xi B_u -A_u B_\xi )\partial_{\bar u} \no \\ &&
- (A_u B_{\bar u} -
A_{\bar u}B_u )\partial_\xi ] 
\er
where now $A$ and $B$ are functions of $u,\bar u ,\xi$.

The algebra of volume-preserving vector fields closes,
\be 
[{\bf v}^{(Y)},{\bf v}^{(\tilde Y)} ] = {\bf v}^{(\tilde{ \tilde Y})}
\ee
\be \label{vec-comm}
\tilde{ \tilde Y}{}^i = (\partial_j Y^i ) \tilde Y^j - 
(\partial_j \tilde Y^i ) Y^j \quad ,
\qquad \partial_i (g\tilde{ \tilde Y}{}^i )=0 .
\ee

Next, we want to investigate a subset of the volume-preserving 
diffeomorphisms obeying an additional condition, because we will find
in the next section that for many models the corresponding Noether currents
are conserved only for this subset (see Eq. (\ref{part3Y3})). Concretely,
the condition defining the subset is
\be \label{d3-Y3}
\partial_3 Y^3 =0.
\ee
This set does not form a subalgebra, as may be checked easily. Indeed,
for $\partial_3 Y^3 = \partial_3 \tilde Y^3 =0$ we find
\be \label{sub-com}
\partial_3 \tilde{\tilde Y}{}^3 = (\partial_1 Y^3 )
(\partial_3 \tilde Y^1 ) +
(\partial_2 Y^3 ) (\partial_3 \tilde Y^2 ) - 
(\partial_1 \tilde Y^3 ) (\partial_3  Y^1 )
-( \partial_2 \tilde Y^3 ) (\partial_3 Y^2 )
\ee
which is nonzero in general. There exist, however, closed subalgebras
within this set. One subalgebra is defined by the condition that
\be \label{Y^3=0}
Y^3 \equiv 0
\ee
because it follows immediately from Eq. (\ref{vec-comm}) that
$Y^3 =0 \, \wedge \, \tilde Y^3 =0 \quad \Rightarrow \quad 
\tilde{\tilde Y}{}^3 =0$.
In terms of the functions $A$ and $B$ there are different possibilities to
fulfill $Y^3 =0$. One may e.g., choose 
\be 
B = B(X^3) \qquad \Rightarrow \partial_1 B = \partial_2 B =0
\ee
which leads to 
\be
Y^1 = h(\partial_2 A ) (\partial_3 B) \, ,\quad 
Y^2 = -h (\partial_1 A) (\partial_3 B) \, ,\quad Y^3 =0.
\ee
The same $Y^i$ are obtained when the transformation $A\to (\partial_3 B) A$, 
$B\to
X^3$ is performed, therefore a general vector field $Y^i$ of this type is
given by 
\be
Y^1 = h(\partial_2 A )  \, ,\quad 
Y^2 = -h (\partial_1 A)  \, ,\quad Y^3 =0.
\ee
Observe that these are precisely the vector fields leading to the Noether
currents (\ref{G-curr}) of Section 2, once the identification $A\to G$ is
made. Therefore, the geometric method of this section indeed provides
an alternative way to find the Noether currents which have been originally
derived from the generalized curvature representation.

An abelian subalgebra of this algebra is obtained by restricting $A$ to
\be \label{abel-SA}
A=A(a,X^3) \, , \quad  B=X^3 \, , \qquad a=(X^1)^2 + (X^2)^2 = u\bar u
\ee 
in which case
\be
Y^1 =2 X^2 h A_{a} \quad ,\quad Y^2 =-2X^1 h A_{a} .
\ee
An apparently different choice with $Y^3=0$ is $A=A(a,X^3)$, $B=B(a,X^3)$, 
but this leads to ($\xi \equiv X^3$, $B_\xi \equiv \partial_\xi B$)
\be
Y^1 = 2 X^2 h (A_a B_\xi -A_\xi B_a)\quad ,\quad Y^2 =-2X^1 h 
(A_a B_\xi -A_\xi B_a)
\ee
and therefore to the same abelian subalgebra (both $Y^1$ and $Y^2$ depending
on one single function of $a$ and $\xi$).

There seems to exist another subalgebra, namely $\partial_3 Y^i=0$ for all
three components of $Y$, which indeed implies $\partial_3 
\tilde{\tilde Y}{}^i =0$, see Eq. (\ref{vec-comm}). However, we have not been
able to find a nontrivial solution to this condition for nontrivial $g$
(i.e., $g_\xi \not= 0$). 

Another set of vector fields obeying $\partial_3 Y^3 =0$ and 
$\partial_i (gY^i) =0$ can be found
for special factorising $g$ of the form
\be 
g=g^{(1)} (X^1,X^2)g^{(2)} (X^3).
\ee
This set is given by
\bdi
Y^1 = g^{-1} (\partial_3 g^{(2)}) H^{(1)} \, ,\quad Y^2 =   
(g)^{-1} (\partial_3 g^{(2)})
H^{(2)} \, , 
\edi
\be
Y^3 =- (g^{(1)})^{-1} (\partial_1 H^{(1)} + \partial_2 H^{(2)}) 
\ee
where $H^{(1)} = H^{(1)} (X^1 ,X^2)$ and $H^{(2)} = 
H^{(2)} (X^1 ,X^2)$ are arbitrary
functions of $X^1$ and $X^2$ only. 
These vector fields lead precisely to the Noether currents 
(\ref{H-curr1}) and (\ref{H-curr2}) of Section 2.

In the special case of $H^{(2)}=0$ the corresponding Clebsch variables are
\be
A = \sqrt{2g^{(2)} H^{(1)}} \cos X^2 \, ,
\quad B= -\sqrt{2g^{(2)} H^{(1)}} \sin X^2
\ee
and for $H^{(1)}=0$ they are
\be
A = \sqrt{2g^{(2)} H^{(1)}} \cos X^1 \, ,
\quad B= \sqrt{2g^{(2)} H^{(1)}} \sin X^1
\ee
whereas they are more complicated in the general case.
This set of vector fields does not form a subalgebra, 
however, for general $H^{(1)}$ and $H^{(2)}$ (i.e. 
$ \partial_3 \tilde{\tilde Y}{}^3 =0
$ does not hold in general). It does form a subalgebra for special
choices of $H^{(1)}$, $H^{(2)}$, like, 
e.g., $H^{(1)} =H^{(1)}(X^2)$ and $H^{(2)} =H^{(2)}(X^1)$, 
or for $H^{(1)} = \partial_2 H$ and $H^{(2)} =-\partial_1 H$. 
But in these special cases $Y^3=0$
and, therefore, they are included in the subalgebra discussed 
above.   

Finally, we give the general expression for 
Noether currents in a relativistic field
theory which correspond to the
vector fields ${\bf v}^{(Y)}$ that generate volume preserving diffeomorphisms
on target space. They are 
\be \label{v-noet-cu}
J^{(Y)}_\mu = Y^u \pi_\mu + Y^{\bar u} \bar \pi_\mu + Y^\xi P_\mu
\ee
where $\pi_\mu =\partial_{u^\mu} {\cal L}$ and $P_\mu = \partial_{\xi^\mu} 
{\cal L}$ are the usual canonical four-momenta, and $u^\mu \equiv
\eta^{\mu\nu}\partial_\nu u$, etc. ($\eta^{\mu\nu}$ is the space-time metric).
The charges $Q^{(Y)}= \int d^3 {\bf r} J^{(Y)}_0$ generate the algebra
of the vector fields ${\bf v}^{(Y)}$ under the Poisson bracket
\be 
\{ u({\bf r}_1) ,\pi_0 ({\bf r}_2) \} =\delta ({\bf r}_1 - {\bf r}_2) \, ,
\quad \{ \xi({\bf r}_1) ,P_0 ({\bf r}_2) \} =\delta ({\bf r}_1 - {\bf r}_2)
\ee
as usual.

\section{Conserved currents in Skyrme type models}

We shall now perform the analysis of conservation laws   
along the lines of what was done for Faddeev--Niemi and related
models in \cite{ASG1}, \cite{ASGW-Ab},
investigating the possibilities of conserved currents for Skyrme and related
models, with special attention to the integrability conditions and the
sectors they define.
First, let us  introduce the abbreviations
\be
a=u\bar u\, ,\quad b=u^\mu \bar u_\mu \, ,\quad c=
(u^\mu \bar u_\mu )^2 - u_\mu^2 \bar u_\nu^2
\ee
\be
d= \xi^\mu \xi_\mu \, ,\quad e = \xi^\mu u_\mu \xi^\nu \bar u_\nu
\ee
such that the quadratic and the quartic part of the Skyrme Lagrangian 
may be re-expressed like
\be
{\cal L}_2 = d + 4 \frac{\sin^2 \xi}{(1+a)^2} b
\ee
and
\be
{\cal L}_4 = \frac{\sin^2 \xi}{(1+a)^2} (bd -e) + \frac{\sin^4 \xi}{(1+a)^4}c.
\ee
Further, we will also study more general Lagrangians ${\cal L} =
{\cal L}(a,b,c,\xi ,d,e)$ with the following canonical four-momenta
\be
\pi_\mu = {\cal L}_{u^\mu} = \bar u_\mu {\cal L}_b +2 (b \bar u_\mu -
\bar u_\nu^2 u_\mu ){\cal L}_c + (\xi^\nu \bar u_\nu) \xi_\mu {\cal L}_e
\ee
\be
P_\mu = {\cal L}_{\xi^\mu} = 2 \xi_\mu {\cal L}_d + ( (\xi^\nu \bar u_\nu)
u_\mu + (\xi^\nu  u_\nu) \bar u_\mu ) {\cal L}_e
\ee
field equations
\be
\partial^\mu \pi_\mu = {\cal L}_u = \bar u {\cal L}_a \, , \quad
\partial^\mu P_\mu = {\cal L}_\xi
\ee
and the following useful identities
\br
u^\mu \pi_\mu &=& b{\cal L}_b +2 c{\cal L}_c + e{\cal L}_e \\
\bar u^\mu \pi_\mu &=& \bar u_\mu^2 {\cal L}_b + (\bar u_\mu \xi^\mu )^2
{\cal L}_e \\
\xi^\mu \pi_\mu &=& (\xi^\mu \bar u_\mu) {\cal L}_b +
2 (b \xi^\mu \bar u_\mu - \bar u_\nu^2 \xi^\mu u_\mu) {\cal L}_c + 
d \xi^\mu \bar u_\mu {\cal L}_e \\
u^\mu P_\mu &=& 2(\xi^\mu u_\mu ) {\cal L}_d + ((\xi^\mu \bar u_\mu ) u_\nu^2 +
b \xi^\mu u_\mu ) {\cal L}_e \\
\xi^\mu P_\mu &=& 2 d {\cal L}_d +2 e{\cal L}_e .
\er 
Now we want to calculate the divergence of the Noether currents
(\ref{v-noet-cu})
\br
\partial^\mu J^{(Y)}_\mu &=& (Y^u_u u^\mu + Y^u_{\bar u} \bar u^\mu
+ Y^u_\xi \xi^\mu )\pi_\mu + 
(Y^{\bar u}_u u^\mu + Y^{\bar u}_{\bar u} \bar u^\mu
+ Y^{\bar u}_\xi \xi^\mu )\bar \pi_\mu + \no \\
&&(Y^\xi_u u^\mu + Y^\xi_{\bar u} \bar u^\mu
+ Y^\xi_\xi \xi^\mu )P_\mu + 
 Y^u \partial^\mu \pi_\mu + Y^{\bar u} \partial^\mu \bar \pi_\mu +
Y^\xi \partial^\mu P_\mu . \no \\
&&
\er
In a first step we want to restrict to the strong integrability conditions
\be \label{strong-int}
u^\mu \xi_\mu =0 \quad ,\quad u^2_\mu =0.
\ee
These are the integrability conditions which follow from the generalized
zero curvature representation for the Skyrme model 
and lead to the infinitely many conserved currents (\ref{G-curr}), 
(\ref{H-curr1}) and (\ref{H-curr2}) in this case, see \cite{FSG1}.
The strong integrability conditions imply
\be
u^\mu \bar \pi_\mu =0 \, , \quad u^\mu P_\mu =0 \, ,\quad \xi^\mu
\pi_\mu =0 ,
\ee
and we find for the current divergence
\be
\left. \partial^\mu J^{(Y)}_\mu \right|_{\rm s} =(Y^u_u + Y^{\bar u}_{\bar u})
u^\mu \pi_\mu +  Y^u {\cal L}_u + Y^{\bar u} {\cal L}_{\bar u} 
+ Y^\xi_\xi \xi^\mu
P_\mu + Y^\xi {\cal L}_\xi ,
\ee
where we used $u^\mu \pi_\mu = \bar u^\mu \bar \pi_\mu$. Next  
we express $Y^i$ like
\be
Y^i =g^{-1}Z^i \equiv hZ^i \qquad \Rightarrow \qquad \partial_i Z^i =0
\ee
and get
\br
\left. \partial^\mu J^{(Y)}_\mu \right|_{\rm s} &=&  Z^u
(h_u u^\mu \pi_\mu + h{\cal L}_u) + Z^{\bar u}
(h_{\bar u} u^\mu \pi_\mu + h{\cal L}_{\bar u} )+ \no \\ &&
Z^\xi (h_\xi u^\mu \pi_\mu +h 
{\cal L}_\xi ) + (hZ^\xi )_\xi (\xi^\mu P_\mu - u^\mu \pi_\mu ) .
\er
Now we assume that $g=g(a,\xi)$ and ${\cal L}={\cal L}(a,\ldots )$ (remember
$a\equiv u\bar u$), which holds in all cases we want to study, and get
\br \label{dJ-s-1} 
\left. \partial^\mu J^{(Y)}_\mu \right|_{\rm s} &=&  (\bar uZ^u
 + uZ^{\bar u})
(h_a u^\mu \pi_\mu + h{\cal L}_a )+  \\ \label{dJ-s-2} &&
Z^\xi (h_\xi u^\mu \pi_\mu + h {\cal L}_\xi ) + \\ \label{dJ-s-3} && 
(hZ^\xi )_\xi (\xi^\mu P_\mu - u^\mu \pi_\mu ) .
\er
If we do not assume any restriction on the Lagrangian, this expression is
zero provided that $Z^\xi =0$ and $(\bar uZ^u + uZ^{\bar u}) =0$. The
latter equation has the general solution $Z^u = iG_{\bar u} =iuG_a$, 
$Z^{\bar u} =-iG_u =-i \bar u G_a$ where $G(a,\xi)$ is an arbitrary function 
of its arguments. In short, the set of conserved currents in the strong 
integrability subsector and for general Lagrangian is given by the
vector fields
\be \label{abel-2}
Y^u  =ih G_{\bar u} =ih  uG_a \, , \quad  Y^{\bar u} =-ihG_u =-ih\bar u G_a
\, ,\quad Y^\xi =0 \, ,\qquad G=G(a,\xi) .
\ee
In terms of the Clebsch variables $A$, $B$, this set is given by
$A=G(a,\xi)$, $B=\xi$, so it is precisely equal to the abelian subalgebra
of Eq. (\ref{abel-SA}).

Next, we want to restrict the possible Lagrangians so that the currents 
remain conserved for a larger class of $Y^i$. The first term,
Eq. (\ref{dJ-s-1}), is zero provided that
\be
(h_a u^\mu \pi_\mu + h{\cal L}_a ) =0,
\ee
or, more explicitly,
\be \label{eq-geom-1}
h_a (b {\cal L}_b +2 c {\cal L}_c + e {\cal L}_e ) +h {\cal L}_a =0.
\ee
This linear first order PDE may be easily solved by the method of
characteristics and has the general solution
\be
{\cal L}(a,b,c,\xi,d,e) = {\cal F}(\frac{b}{h},\frac{c}{h^2},\xi ,d,
\frac{e}{h})
\ee
where ${\cal F}$ is an arbitrary function of its arguments. Therefore,
Eq. (\ref{eq-geom-1}) fixes the dependence on $a$ in terms of the dependence 
on $h$ (i.e., $g=h^{-1}$). Eq. (\ref{dJ-s-2}) is zero if
\be
(h_\xi u^\mu \pi_\mu + h{\cal L}_\xi ) =0,
\ee
or
\be \label{eq-geom-2}
h_\xi (b {\cal L}_b +2 c {\cal L}_c + e {\cal L}_e ) +h {\cal L}_\xi =0
\ee
with the general solution
\be
{\cal L}(a,b,c,\xi,d,e) = {\cal F}(a,\frac{b}{h},\frac{c}{h^2} ,d,
\frac{e}{h}) .
\ee
A general solution to both equations is, therefore,
\be \label{sol-geom-12}
{\cal L}(a,b,c,\xi,d,e) = {\cal F}(\frac{b}{h},\frac{c}{h^2} ,d,
\frac{e}{h})
\ee
and fixes the dependence both on $a$ and on $\xi$ in terms of a dependence
on $h$.
The Skyrme model is in this class with
\be
g=h^{-1} = 4 \frac{\sin^2 \xi}{(1+a)^2} .
\ee
Finally, expression (\ref{dJ-s-3}) is zero if the Lagrangian obeys
\be
(\xi^\mu P_\mu - u^\mu \pi_\mu )=0
\ee
or, more explicitly,
\be \label{weight-eq}
b {\cal L}_b +2 c {\cal L}_c -2d{\cal L}_d - e {\cal L}_e =0.
\ee
There exist Lagrangians which obey this additional condition, which may
contain terms  like, e.g., $ (b/h)^2 d$ or $(c/h^2)d$, etc. 
In fact, if we introduce a ``weight number'' which associates weight
$+1$ with each power of the derivative $u_\mu$ and $\bar u_\mu$, and a weight
$-2$ with each power of the 
derivative $\xi_\mu$, then Eq. (\ref{weight-eq}) just
requires that the total weight of each term in the Lagrangian is zero, i.e.
\be \label{weight}
{\cal W} \equiv {\rm pow}(u_\mu )+{\rm pow}(\bar u_\mu ) 
- 2 {\rm pow}(\xi_\mu )
=0
\ee
(where e.g., ${\cal W}(b)=2$ or $ {\cal W}(e)=-2$ or $ {\cal W}(d)=-4$).

The Skyrme Lagrangian, however, does not obey Eq. (\ref{weight-eq}). 
Therefore,
in this case Expression (\ref{dJ-s-3}) is zero only provided that
\be \label{part3Y3}
(hZ^\xi )_\xi \equiv Y^\xi_\xi \equiv \partial_3 Y^3 =0,
\ee
that is, exactly  condition (\ref{d3-Y3}) of the last section.
Therefore, a general Noether current $J^{(y)}_\mu$ is conserved in the
strong integrability subsector of the Skyrme model provided that $Y$ 
obeys Eq. (\ref{part3Y3}). The currents (\ref{G-curr}), (\ref{H-curr1}) 
and (\ref{H-curr2}) belong to this class, but there may exist more
generators $Y$ obeying condition (\ref{part3Y3}).

Next, let us relax the strong integrability conditions. We still require that
\be
u^\mu \xi_\mu =0 
\ee
but now $u_\mu^2$ need not be zero. This implies
\be
u^\mu \bar \pi_\mu =u_\mu^2 {\cal L}_b 
\, , \quad u^\mu P_\mu =0 \, ,\quad \xi^\mu
\pi_\mu =0
\ee
and leads to the current divergence
\be
\left. \partial^\mu J^{(Y)}_\mu \right|_{\rm w} =
\left. \partial^\mu J^{(Y)}_\mu \right|_{\rm s} + ( Y^u_{\bar u} 
\bar u_\mu^2  + Y^{\bar u}_u u_\mu^2 ) {\cal L}_b .
\ee
This divergence is zero provided that, in addition to 
$\left. \partial^\mu J^{(Y)}_\mu \right|_{\rm s} =0$,  either
$ {\cal L}_b =0$, i.e., the Lagrangian does not depend on $b$, or 
$Y$ is again restricted to the abelian subalgebra (\ref{abel-2}). Then
the above divergence can be reexpressed like
\be
\left. \partial^\mu J^{(Y)}_\mu \right|_{\rm w} =
\left. \partial^\mu J^{(Y)}_\mu \right|_{\rm s} + [\partial_a (hG_a)] ( 
u^2 \bar u_\mu^2  - \bar u^2 u_\mu^2 ) {\cal L}_b
\ee
and can be made zero by imposing the weak integrability condition
\be \label{weak-1}
u^2 \bar u_\mu^2  - \bar u^2 u_\mu^2 =0
\ee
in addition to $u^\mu \xi_\mu =0$. Therefore, the currents $J^{(Y)}_\mu$
are conserved for $Y$ belonging to the abelian subalgeba (\ref{abel-2}) 
if the weak integrability conditions 
\be \label{weak-2}
u^2 \bar u_\mu^2  - \bar u^2 u_\mu^2 =0 \, ,\quad 
u^\mu \xi_\mu =0
\ee 
are fulfilled. This is true for arbitrary Lagrangians, because 
$\left. \partial^\mu J^{(Y)}_\mu \right|_{\rm s}$ is identically zero
for $Y$ belonging to the abelian subalgebra (\ref{abel-2}).
Therefore, these conserved currents exist in the weak integrable subsector
of the Skyrme model.

The weak integrability conditions (\ref{weak-2}) have a nice geometrical
interpretation in terms of the real functions $(\xi ,\Sigma ,\Lambda )$ 
where $u =\exp (\Sigma + i \Lambda )$, see (\cite{ASGW-SkCPn}). 
They are then equivalent to the
perpendicularity conditions
\be
\xi^\mu \Sigma_\mu =0 \, ,\quad \xi^\mu \Lambda_\mu =0 \, , \quad
\Sigma^\mu \Lambda_\mu =0.
\ee
Specifically, for time-independent field configurations these are just the
conditions that $(\xi ,\Sigma ,\Lambda )$ form a set of perpendicular
curvilinear coordinates in base space $\RR^3$.

Finally, let us calculate the full divergence. We get
\br
\partial^\mu J^{(Y)}_\mu &=& \left. \partial^\mu J^{(Y)}_\mu \right|_{\rm w}
+ \{ (u^\mu \xi_\mu )^2 {\cal L}_e Y^{\bar u} _u + \label{full-1}  \\
&& u^\mu \xi_\mu [({\cal L}_b +2b {\cal L}_c +d{\cal L}_e ) Y^{\bar u}_\xi
+ (2{\cal L}_d +b{\cal L}_e )Y^\xi_u ] \label{full-2} \\
&&+ u_\nu^2 \bar u^\mu \xi_\mu (2{\cal L}_c Y^{\bar u}_\xi - {\cal L}_e
Y^\xi_u ) \, + \, {\rm h.c.} \} \label{full-3}
\er
which can be made equal to zero in different ways. 

If no constraints (integrability conditions) are imposed, obviously
only the Noether
currents of the target space symmetries remain. For a completely generic
Lagrangian the only remaining conserved current is the one with $Y$ given by
\be
Y^u =iu \, , \quad Y^{\bar u} =-i\bar u \, ,\quad Y^\xi =0
\ee
which corresponds to the symmetry $u\to e^{i\alpha }u$ and is always present
for Lagrangians ${\cal L}={\cal L}(a,b,c,\xi ,d ,e)$ by construction.
If the Lagrangian obeys the condition (\ref{sol-geom-12}), like is the case,
e.g., in the Skyrme model, then still
there exist only finitely many conserved currents and, moreover, the
corresponding charges are now the Noether charges of the isometries of the 
(target space) metric
\be \label{tar-met}
ds^2 = d\xi^2 +g dud\bar u .
\ee
But there also 
exist Lagrangians with infinitely many symmetries. Indeed, let us
restrict to a subalgebra of the abelian subalgebra (\ref{abel-2}) such
that $\partial_\xi Y^u =0$ in addition to $Y^\xi =0$. This is achieved by 
setting
\be \label{abel-3}
Y^u  =iu\tilde G_a \, ,\quad Y^{\bar u} =-i\bar u \tilde G_a 
\, ,\quad Y^\xi =0 \, ,\qquad \tilde G=\tilde G(a) \, ,\quad 
\tilde G_\xi \equiv 0 .
\ee
These vector fields obey 
$\partial_i (gY^i)=\partial_u (gY^u) + \partial_{\bar u} (g
Y^{\bar u})=0$ for $g=g(a,\xi)$, 
as may be checked easily, and are, therefore,  volume
preserving diffeomorphisms forming a subalgebra of the abelian subalgebra
(\ref{abel-2}). In the above current divergence, the second and third line
(Eqs. (\ref{full-2}) and (\ref{full-3}))
are zero for  $Y$ of this type, and only the first line, Eq. (\ref{full-1}),
remains. The second term of the first line vanishes if we assume
${\cal L}_e =0$, i.e., ${\cal L} = {\cal L}(a,b,c,\xi ,d)$. 
Further,  
$\left. J^{(Y)}_\mu \right|_{\rm s}$ is
automatically conserved. It remains to investigate the additional term
in $\left. J^{(Y)}_\mu \right|_{\rm w}$. 
For ${\cal L}_b=0$ this additional term
is zero, and it follows that a field theory with a Lagrangian
${\cal L}(a,c,\xi ,d)$ has infinitely many conserved currents 
$J^{(Y)}_\mu $ with $Y$ given by (\ref{abel-3}) for the full model
(i.e., without additional integrability conditions). It has, therefore,
infinitely many target space Noether symmetries.

Alternatively, 
an arbitrary model ${\cal L}(a,b,c,\xi ,d)$ has infinitely many
conserved currents $J^{(Y)}_\mu $ with $Y$ given by (\ref{abel-3}) in the
submodel defined by the weak integrability condition (\ref{weak-1}).

As we have seen,
there are quite many possibilities for having conserved currents
for different models (i.e., different Lagrangians) or their submodels
defined by some additional integrability conditions.
Therefore, we summarize our results
in the Tables 1--3. Observe that there exist infinitely many conserved
currents in all cases of Table 1 and Table 2. In Table 3 there are only
finitely many conserved currents in case a) and b), whereas there exist
infinitely many conservation laws in the remaining cases. 

The Skyrme model without the pion mass term
corresponds to case b) of Table 1 (for the strong
integrability conditions (\ref{strong-int})), to case a) of Table 2
(for the weak integrability conditions (\ref{weak-2}), or to case b) of
Table 3 (the isometries of the target space $S^3$).
The Skyrme model with pion mass term corresponds to case a) of Table 1
and to case a) of Table 2, that is, it has the same conserved currents
for the strong and for the weak intergrability conditions. Further, it
corresponds to case a) of Table 3 where it has, however, three conserved
currents (corresponding to three target space symmetries) rather than
just one in the most generic case (i.e., the Skyrme model Lagrangian with
pion mass term is still somewhat ``special'' and has an enhanced symmetry).

Recently, another class of models has been studied, and infinitely many
soliton solutions have been found analytically (see Ref. \cite{ASGW-SkCPn}), 
for the class of Lagrangians
\be
{\cal L}= - f^{(1)}(a) f^{(2)}(\xi) c^\frac{3}{4} + d^\frac{3}{2},      
\ee
where the non-integer power of the kinetic terms has been chosen carefully 
in order
to avoid the Derrick scaling argument against the existence of soliton
solutions. Further, $f^{(1)}$ and $f^{(2)}$ are arbitrary functions of their
arguments. Lagrangians of this type belong to case c) of Table 3 and
have, therefore, infinitely many target space symmetries and infinitely
many conserved charges.

\section{Conclusions}
We have used geometric structures of the target space to
find and classify the conservation laws in a large class of Skyrme
type models. More precisely, we have analysed under which conditions there
exist infinitely many conservation laws either in the full theories or
in submodels defined by additional integrability conditions.
It turned out that the conserved charges belong to certain subsets
of the Noether charges of volume-preserving target space
diffeomorphisms in all cases.
These conservation laws should be  helpful for the further study of these 
models, e.g., for 
finding solutions, either exact or numerical. In more general terms, 
the integrability conditions could serve as an alternative to
the BPS conditions, which are missing in the Skyrme model - in analogy to
the CP$^n$ models, where the strong integrable subsector is precisely 
equivalent to the BPS sector, whereas the weak integrable subsector also
allows for non-BPS solutions, see \cite{ASGW-SkCPn}.

Specifically, for the Skyrme model we were able to re-derive the
results of \cite{FSG1} and of  \cite{ASGW-SkCPn} on the strong and weak
integrable subsectors of the Skyrme theory, and to put these results into
a more geometric context. All the conserved currents in the strong integrable
subsector of the Skyrme model found in \cite{FSG1} belong to the subset
of the volume preserving diffeomorphisms obeying $\partial_3 Y^3 =0$, see
e.g. case b) in Table 1. It is however possible that this set
is not exhausted by the currents given explicitly in \cite{FSG1} (see
Eqs. (\ref{G-curr}), (\ref{H-curr1}),  (\ref{H-curr2})), that is, there might
exist more currents in this subset.
Interestingly, this subset does not form a subalgebra.
The conserved charges in the weak integrable subsector of the Skyrme model
found in \cite{ASGW-SkCPn} form an abelian subalgebra of the Lie algebra
of volume preserving target space diffeomorphisms, which is given e.g. in
case a) of Table 2. The fact that the conserved charges do form a Lie
subalgebra for the weak integrability conditions but not for the strong
integrability conditions makes the former ones appear somewhat more natural.

Further, the results of \cite{ASGW-Ab} for a two-dimensional
target space may be recovered easily from the results presented here by simply
assuming that nothing depends on the third target space coordinate $X^3
\equiv \xi$ and by setting equal to zero the corresponding vector component,
i.e., $Y^3 =0$. 
Finally, let us remark that
our results could also shed more light on the original zero curvature
construction of \cite{AFSG},
and  might, for instance, help in finding the  appropiate 2 form  or even more
general connections for this construction. This problem 
is under investiagtion.

\begin{table}

\begin{tabular}{|rl|}
\hline
\multicolumn{2}{|c|}{Integrability conditions 
$u_\mu^2 =0$ and $u^\mu \xi_\mu =0$.} \\
\hline 
 a) & no condition on ${\cal L}$.  \\
& $Y$ forms the abelian subalgebra (for $G=G(a,\xi)$): \\
& $Y^u = ihuG_a$, \, $Y^{\bar u} =-ih\bar u G_a$, \, $Y^\xi =0$. \\
\hline
b) & ${\cal L}= {\cal F}(\frac{b}{h},\frac{c}{h^2},d ,\frac{e}{h})$, 
see Eq. (\ref{sol-geom-12}) \\
& $Y$ form the subset $Y^\xi_\xi =0$ (is {\em not}  a subalgebra). \\
\hline
c) &  ${\cal W}({\cal L})   =0$, see Eq. (\ref{weight}). \\
 & no further condition on $Y$.   \\
\hline
\end{tabular} 

\caption{Conserved currents $J^{(Y)}_\mu$ 
for the strong integrability conditions.  \newline
The vector fields
$Y$ always generate volume-preserving diffeomorphisms, i.e., they
obey $\partial_i (gY^i)=0$. \newline
A general Lagrangian is of the form ${\cal L}= {\cal L} (a,b,c,\xi ,d,e)$.
}

\end{table}

\begin{table}

\begin{tabular}{|rl|}
\hline
\multicolumn{2}{|c|}{Integrability conditions $u^\mu \xi_\mu =0$.} \\
\hline 
 a) & no condition on ${\cal L}$; or ${\cal L}= 
{\cal F}(\frac{b}{h},\frac{c}{h^2},d ,\frac{e}{h})$.  \\
& $Y$ forms the abelian subalgebra (for $G=G(a,\xi)$): \\
& $Y^u = ihuG_a$, \, $Y^{\bar u} =-ih\bar u G_a$, \, $Y^\xi =0$. \\
& {\em And} the further integrability condition 
 $u^2 \bar u_\mu^2 - \bar u^2 u_\mu^2 $ holds. \\
\hline
b) & ${\cal L}_b =0$. \\
& $Y$ forms the abelian subalgebra (for $G=G(a,\xi)$): \\
& $Y^u = ihuG_a$, \, $Y^{\bar u} =-ih\bar u G_a$, \, $Y^\xi =0$. \\
\hline
c) & ${\cal L}_b =0$ and ${\cal L}= 
{\cal F}(\frac{b}{h},\frac{c}{h^2},d ,\frac{e}{h})$. \\
& $Y$ form the subset $Y^\xi_\xi =0$ (is {\em not}  a subalgebra). \\
\hline
d) &   ${\cal L}_b =0$ and ${\cal L}= 
{\cal F}(\frac{b}{h},\frac{c}{h^2},d ,\frac{e}{h})$
and ${\cal W}({\cal L})  =0$. \\
 & no further condition on $Y$.   \\
\hline
\end{tabular} 

\caption{Conserved currents $J^{(Y)}_\mu$ 
for the weak integrability conditions.  \newline
The vector fields 
$Y$ always generate volume-preserving diffeomorphisms, i.e., they
obey $\partial_i (gY^i)=0$. \newline
A general Lagrangian is of the form ${\cal L}= {\cal L} (a,b,c,\xi ,d,e)$.
}

\end{table}

\begin{table}

\begin{tabular}{|rl|}
\hline
\multicolumn{2}{|c|}{No integrability conditions. } \\
\hline 
 a) & no condition on ${\cal L}$.  \\
& Generically there exists only one vector field $Y$: \\
& $Y^u = iu$, \, $Y^{\bar u} =-i\bar u $, \, $Y^\xi =0$. \\
\hline
b) & ${\cal L}= 
{\cal F}(\frac{b}{h},\frac{c}{h^2},d ,\frac{e}{h})$. \\
& There exist finitely many $Y$ generating the \\
& target space isometries for the metric of Eq. (\ref{tar-met}). \\
\hline
c) & ${\cal L}_b =0$ and ${\cal L}_e =0$.  \\
& $Y$ form the abelian subalgebra (see Eq. (\ref{abel-3}), where
$\tilde G=\tilde G(a)$): \\
& $Y^u  =iu\tilde G_a$, \, $ Y^{\bar u} =-i\bar u \tilde G_a $,
\, $ Y^\xi =0$. \\
\hline
d) &   ${\cal L}_b =0$ and ${\cal L}_e =0$.  \\
& $Y$ forms the abelian subalgebra (for $G=G(a,\xi)$): \\
& $Y^u = ihuG_a$, \, $Y^{\bar u} =-ih\bar u G_a$, \, $Y^\xi =0$. \\
& {\em And} the further integrability condition 
 $u^2 \bar u_\mu^2 - \bar u^2 u_\mu^2 $ holds. \\
\hline
e) & ${\cal L}_e =0$.  \\
& $Y$ form the abelian subalgebra (see Eq. (\ref{abel-3}), where
$\tilde G=\tilde G(a)$): \\
& $Y^u  =iu\tilde G_a$, \, $ Y^{\bar u} =-i\bar u \tilde G_a $,
\, $ Y^\xi =0$. \\
& {\em And} the further integrability condition 
 $u^2 \bar u_\mu^2 - \bar u^2 u_\mu^2 $ holds. \\

\hline
\end{tabular} 

\caption{Conserved currents $J^{(Y)}_\mu$ 
without further integrability conditions.  \newline
The vector fields 
$Y$ always generate volume-preserving diffeomorphisms, i.e., they
obey $\partial_i (gY^i)=0$. \newline
A general Lagrangian is of the form ${\cal L}= {\cal L} (a,b,c,\xi ,d,e)$.
}

\end{table}

\section*{Acknowledgements}
C. A. and J.S.-G. thank MCyT (Spain) and FEDER (FPA2005-01963),
and Incentivos from Xunta de Galicia. A.W. gratefully acknowledges
Departamento de Fisica de Particulas, Universidad de Santiago for
hospitality.
Further, C. A. acknowledges support from the
Austrian START award project FWF-Y-137-TEC and from the FWF
project P161 05 NO 5 of N.J. Mauser.

\end{document}